# Improving Customer Service in Healthcare with CRM 2.0

Mohammad Nabil Almunawar[#1] and Muhammad Anshari[#2]

*Faculty of Business, Economic & Policy Studies*
*Universiti Brunei Darussalam*
*Brunei Darussalam*

[1]nabil.almunawar@ubd.edu.bn
[2]anshari@yahoo.com

*Abstract*— The Healthcare industry is undergoing a paradigm shift from healthcare institution-centred care to a citizen-centred care that emphasises on continuity of care from prevention to rehabilitation. The recent development of Information and Communication Technology (ICT), especially the Internet and its related technologies has become the main driver of the paradigm shift. Managing relationship with customers (patients) is becoming more important in the new paradigm. The paper discusses Customer Relationship Management (CRM) in healthcare and proposes a Social CRM or CRM 2.0 model to take advantage of the multi-way relationships created by Web 2.0 and its widespread use in improving customer services for mutual benefits between healthcare providers and their customers.

*Keywords*— Health Customer Service, CRM, CRM 2.0, Social CRM, Value Creation

## I. INTRODUCTION

One of the most interesting aspects in healthcare management is how to manage the relationship between a healthcare provider and its customers (patients) in order to create a greater mutual understanding, trust, and patient involvement in decision making. A good relationship between a healthcare provider and its customers will lead to improve customers' satisfaction, which in turn make them loyal customers [26].

A good relationship between a healthcare provider and its customers does not only improve customer's satisfaction, but also helps in fostering effective communications between them, which may help to improve their health and health-related quality life and more effective in chronic disease management [1]. On the other hand, failure in managing the relationship will create dissatisfaction of customers, which may lead to distrust towards the system. In addition, bad or unmanaged relationship with customers will make them feel alienated during a treatment; in summary, the business sustainability of the healthcare provider will be threatened by sour relationship with its customers. Therefore, a good relationship between a healthcare provider and its customers is crucial and the relationship must be managed effectively to sustain the business.

The roles of Customer Relationship Management (CRM) in managing customers and improving services to them have been well-recorded in business literatures. Many corporations these days, including those in the healthcare industry use CRM as a tool to serve customers better. Corporations take advantage of the recent development in Information and Communication Technology (ICT), especially the Internet related technology; embracing e-business. CRM is an integral part of e-business architecture, meaning that e-Business without CRM is not complete.

CRM can be viewed as strategy to retain existing customers and attract new ones. Customer retention is important for growth and sustainability of the business. CRM can also be used to extend other services or products to the customers. In the healthcare environment, healthcare providers are challenged not only to retain existing customers but also to acquire potential customers for the healthcare services, retaining them to use the services, and extending various services in the future. With the growing competition among healthcare providers, managing the customer relationship and providing better services through CRM is a strategy that needs to be carefully planned.

Unfortunately, many see CRM is merely a technology for improving customer service which may lead to a failure when implemented. CRM initiatives must be seen as a strategy for significant improvement in services by solidifying satisfaction, loyalty and advocacy through information and communication technology. As such, matters pertaining related to people such as customer behaviour, culture transformation, personal agendas, and new interactions between individuals and group must be incorporated in CRM initiatives. Therefore, an organization needs to understand that behaviour and expectations of customers which continue to change overtime. Consequently CRM must address the dynamic nature of customers' needs and adjustment strategies embedded in CRM are required.

With the recent development in ICT, especially the emerging Web 2.0 technology such as social networks (e.g. Facebook, Twitter), blogs, wikis, and video sharing (Youtube), the capability of current CRM technology can be enhanced further. Greenberg [14] proposes the new CRM with the Web 2.0 technology as Social CRM or CRM 2.0. The new CRM, which is still in its infancy, has all CRM capabilities with some additional capabilities in term of fast interactivity and empowerment of customers.

The objective of this paper is to lay the foundation for developing a system for managing customer relationships using Web 2.0 technology in healthcare industry. A new model is proposed based on Web 2.0 capabilities and value creation in healthcare industry.

In this paper we propose a new CRM model that accommodates Web 2.0 technology. Although our model is fine-tuned to manage customer relationship in healthcare industry, the model can be modified easily for other industries. Communications in healthcare industry is intense; we argue that our model will improve the relationship between healthcare providers and their customers for mutual benefit. This paper is organised as follows: the next section will discuss with more detail on Social CRM or CRM 2.0, section 3 will explore value creation in healthcare industry, our model will be discussed in section 4, section 5 is the future trend and Section 6 is the conclusion.

## II. LITERATURE REVIEW

Gartner [12] defined CRM as a broad term and widely-implemented strategy for managing interactions with customers which involves using technology to organize, automate, and synchronize business processes—principally customer service, marketing, and sales activities. The overall goals are to find, attract, and win new customers, nurture and retain those the company already has, attract former customers back into the fold, and reduce the costs of marketing and customer service.

Greenberg [14] stated that CRM is a philosophy and a business strategy supported by a system and a technology designed to improve human interactions in a business environment. Furthermore, it is an operational and transactional approach to customer management that is focused around the customer facing departments, sales, marketing and customer service. Furthermore, the early CRM initiatives was the process for modification, culture change, technology and automation through use of data to support the management of customers so it can meet a business value of corporate objectives such as increase in revenue, higher margins, increase in selling time, campaign effectiveness, reduction in call queuing time, etc.

CRM strategy must be aligned to the organization's mission and objectives in order to bring about a sustained performance of business objectives and effective customer relationships. The organization must adopt customer's perspective and work on developing a comprehensive planning write up and specific business objectives. The strategies should be laid down in such a way so that they provide benefits to the company and customers, shorter cycle times, greater customer involvement in service development and reduce operation costs by redesigning business process that eliminates work which does not add value to customers [30].

Yina [32] examined healthcare providers in adopting CRM as a strategy in building trust to their patients as well as helping patients to avoid feel alienated in the healthcare environment and at the same time improving the service quality and efficiency of healthcare [6]. With the Web technology, CRM also affords healthcare providers the ability to extend services beyond its traditional practices, and it provides a competitive advantage environment for a healthcare provider to achieve a complex patient care goal. CRM enables a healthcare provider to capture essential patient (customer) information to be utilized effectively, especially in integrating the patients' information in a system to promote superb service.

Many studies also have reported that many organizations have failed to adopt CRM as a strategy. The main reasons for high rate of failure when implementing CRM are: 1) related to people's behaviour and culture [13], and 2) CRM is viewed as a merely technology, not as a long-term strategy [22]. These notable failures encourage further studies on the implementation issue especially in dealing with change of management. Organizational change is always expected in CRM initiative [14]. Hence, there is a possible area to explore further CRM initiative that focuses on people, technology and culture.

There is also research examining the development of CRM within healthcare environment. There are many challenges in adopting CRM for healthcare organizations. Due to the complexity of the business nature in healthcare, there are many issues dealing with patients that must be considered. A healthcare is undergoing a paradigm shift, from 'Industrial Age Medicine to Information Age Healthcare' [28]. This 'paradigm shift' is reshaping health systems [15]. It is also transforming the healthcare-patient relationship [3]. For example, World Wide Web has changed the way the public engages with health information [25]. According to Pew Internet and American Life Project, large shares of Internet users say that they will first use Internet when they need Information about healthcare [24]. People are beginning to use Internet resources for research on the health information and services that they are interested in using.

Batista [4] studied about the Internet in becoming a crucial medium in supporting CRM. Indeed, the Web technology is a powerful channel available for organizations to utilise to enhance interactions and strengthen relationship with customers.

Social CRM or CRM 2.0 is based on the Web 2.0. The Web 2.0 could be used as enablers in creating close and long term relationships between an organization with its customers [2]. Greenberg [14] defined Social CRM as a philosophy and a business strategy, supported by a technology platform, business rules, processes, and social characteristics, designed to engage the customer in a collaborative conversation in order to provide mutually beneficial value in a trusted and transparent business environment. It's the company's response to the customer's ownership of the conversation.

The term of Social CRM and CRM 2.0 is used interchangeably. Both share new special capabilities of social media and social networks that provide powerful new approaches to surpass traditional CRM.

Cipriani [11] described the fundamental changes CRM 2.0 is introducing to the current, traditional CRM in terms of landscape. The most significant feature of CRM 2.0 is the network among customers and healthcare providers. This network creates value of network such as multi-ways communications and sharing of experience and knowledge. Table 1 summarizes the difference of CRM 2.0 from CRM 1.0 based on type of relationship, connection, and how value generated. Relationship type in CRM 1.0 focuses on the individual relationship; Customer to Customer or Customer to Business whereas in CRM 2.0 offers the collaborative relationship and engage a more complex relationship network. Connection type in CRM 1.0 is a limited view of the customer which adversely affects the less informed customer; on the other hand, CRM 2.0 enables multiple connections which allow customers to be more understanding and knowledgeable. In terms of value creation, CRM 1.0 is constricted to targeted messages, whereas CRM 2.0 offers a more diverse value creation from informal conversation of customers within social networks.

TABLE I
COMPARISON CRM 1.0 AND CRM 2.0

| Type | CRM 1.0 | CRM 2.0 |
|---|---|---|
| **Relationship** | Focus on individual relationship (C2C, C2B) | Focus on collaborative relationship (engaging a more complex relationship network) |
| **Connection** | Limited view of the customer & his community preferences, habits, etc | Multiple connections allow better understanding of the customer and his community |
| **Generated Value** | Targeted messages generate value | Conversation generates value |

Source: Cipriani [11]

## III. VALUE CREATION IN HEALTHCARE INDUSTRY

Web 2.0 services help people and organizations to create customer values and to build customer relationship through collaboration and social networking [27]. It is important to examine each business process as a layer of value to the service. Patients place a value on these services according to quality of outcome, quality of service, and price. The value of each layer depends on how well they are performed. When a healthcare provider cannot achieve its strategic objectives, it needs to reengineer its activities to fit business processes with strategy. If the business processes do not fit the strategy, it will diminish the value. For example, the value of a health education is reduced by a delay respond of patient's query or poor communication skills. The value of service is reduced by a poor schedule of physician.

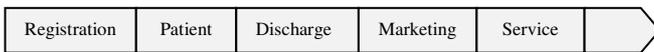

Fig. 2 Healthcare Value Chain [21]

Fig. 2 above is originated from Porter's Value Chain analysis. Each activity adds value to the customer. These are registration, patient care, discharge, marketing, and service. For example, the value on service depends on how well they educate the patient.

Porter [24] proposed value chain framework for the analysis of organizational level competitive strengths and weaknesses. It is a method for decomposing the firm into strategically important activities and understanding their impact on cost and value. This model was originally developed for manufacturing organizations which have complex activities in product design, manufacture, and assembly. However, according to Porter [24], the overall value creating logic of the value chain with its generic categories activities is valid in all industries. Figure 3 shows the scope of CRM within the healthcare value chain based on Porter's model. CRM tools and strategies will be implemented in the marketing and service layer. The healthcare organization should perform re-engineering process to adapt their CRM strategy and tool in order to acquire potential customer coming for the service [1].

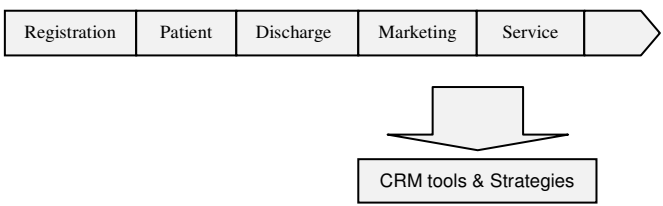

Fig. 3 Healthcare Value Chain with CRM Scope

Further research conducted by Stabell and Fjeldstad [29] refined the three distinct generic value configuration models (value chain, value shop, and value network) required to understand and analyze firm-level value creation logic across broad range of industries. With the identification of alternative value creation technologies, value chain analysis is sharpened a value configuration analysis to gain competitive advantage. The study revealed that implementing the model have experienced serious problems and less suitable in a numbers of service industries. Based on Thompson's [30] typology of long linked, intensive and mediating technologies, Stabell and Fjeldstad suggested that the value chain models the activities of long-linked technology, while the value shop models firm where value is created by mobilizing resources and activities to resolve a particular customer problem, and the value network models firms that create value by facilitating a network relationship between their customers using a mediating technology. Healthcare, schools & universities, and consulting firms are examples of value shop that rely on an intensive technology.

Figure 4 is the generic value shop diagram to the general practitioners. The medical consultation shop appears to be diagnosis-focused shop. Treatment plans follow after the diagnosis. The cyclic nature of the activities set is captured by circular layout of the primary activity categories. The five generic categories of primary value shop activities; Problem-finding and acquisition – activities associated with the recording, reviewing, and formulating of the problem to be solved and choosing the overall approach to solving the problem, problem finding – activities associated with generating and evaluating alternative solutions, choice – activities associated with choosing among alternative problem solutions, execution – activities associated with communicating, organizing, and implementing the chosen solution, and control & evaluation – activities associated with measuring and evaluating to what extent implementation has solved the initial problem statement.

Problem finding & acquisition have much in common with marketing in the value chain; the customer owns the problem in certain cases such as health service and education. Consider marketing, defining client's problem is also client acquisition. Marketing is largely relationship management [8, 9].

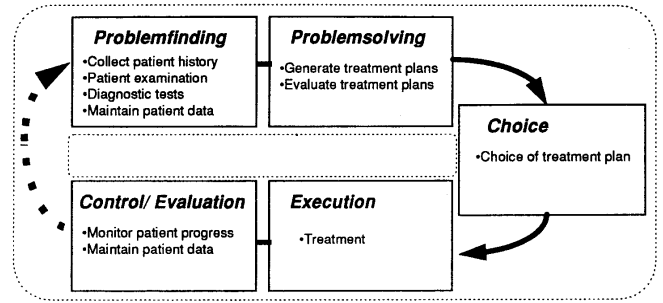

Fig. 4 The Value Shop Diagram for a General Practitioner [29]

With the Web 2.0 in which customers are connected and can easily communicate and share experience as well as communicate with medical professionals, an important value of network is generated. This value will be inherent in our CRM 2.0 that we propose.

## IV. METHODOLOGY

Based on the existing literature on CRM, healthcare industry and value creation we developed a preliminary model of CRM 2.0 for healthcare industry. We derive features of the model and conducted a survey to examine whether features offered by the model concur with features needed by customers. The model is refined based on the survey and will be discussed in the next section.

The survey was conducted at major hospitals, clinics, or home care centres on Brunei Darussalam starting from February to March, 2011. The features of our CRM 2.0 model were measured by descriptive analysis. We use purposive sampling methods in which respondents were intentionally selected from patients, patient's family, or medical staffs from hospitals, clinics, and homecare centre across the country. Majority of the respondents are local Bruneians, who frequently visit health practitioners. The respondents range from twenty years old (or younger) to above 51 years old. Therefore, they represent a fair share of the general public. There were 366 respondents participating for the survey and conducted from February to March 2011.

## V. THE CRM 2.0 MODEL

The model operates in the area of healthcare organization–patient relationships inclusive with social network interaction, and how they can possibly share information to achieve health outcomes. Figure 5 is a proposed model of Social CRM in a healthcare environment. It offers a starting point for identifying possible theoretical mechanisms that might account for ways in which Social CRM provides one-stop service for building relationships between a healthcare organization, patients and community at large.

The framework is developed from Enterprise Social Networks, Internal Social Networks, Listening tool interfaces, Social CRM systems within healthcare provider, and healthcare value configuration (value chain and value shop).

Social Networks refers to any Web 2.0 technology that a patient or his/her families may join. It differentiates two social networks linkages to the patient or his/her family; they are Enterprises Social Networks and Internal Social Networks. The Enterprises Social Networks refers to external and popular Web 2.0 applications such as Facebook, Twitter, LinkedIn, MySpace, Friendster etc. which patients may belong to for interaction. The dashed line connecting enterprises social networks and CRM systems means that none of those networks have control over the others directly, but constructive conversation and information from enterprises social networks should be captured for creating strategy, innovation, better service and at the same time responds accurately.

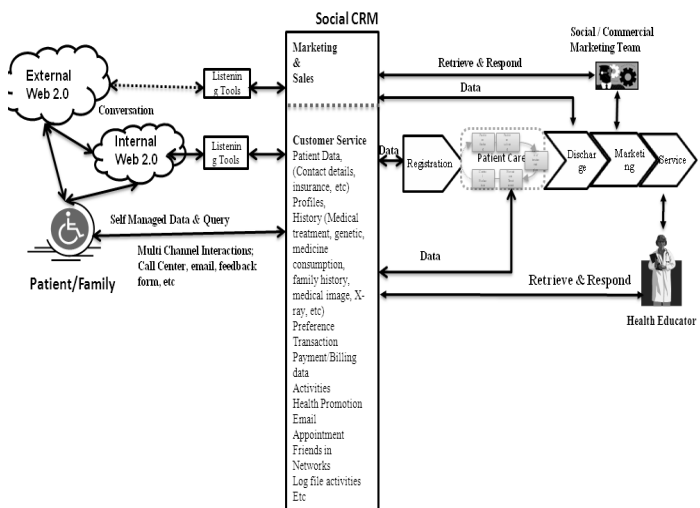

Fig. 5 Proposed Model of Social CRM in Healthcare

Further, the model has Internal Social Networks that is operated, managed and maintained within the healthcare's infrastructure. This is more targeted to internal patients/families within the healthcare to have conversation between patients/family within the same interest or health problem/ illness. For example, patients with diabetes would be motivated to share his/her experiences, learning and knowledge with other diabetic patients. Since the patient/family generates the contents of the Web, it can act as a useful learning center for others, not only promoting health among each other, but also it can be the best place for group support and sharing experiences related to all issues such as: how the treatment is being done by the healthcare, cost, what insurance is accepted by healthcare, food services and nutrition provided etc. Therefore, the growth of this generic group will depend on the need of patients in that healthcare. For instant, N1 is internal social networks for Diabetes, N2 is for Cancer, N3 is for heart disease, and so on. Creating Internal Social Networks is part of the strategy to isolate problems into small space or putting more focus on the local problem so it can be easily monitored and solved. Moreover, this strategy will promote customer loyalty towards the healthcare.

In general, the aim to put together linkage of internal and external social networks are to engage patients and export ideas, foster innovations of new services, quick response/feedback for existing service and technologies from people inside and outside organization. Both provide a range of roles for patient or his/her family. The relationships can create emotional support, substantial aid and service, influence, advice and information that a person can use to deal with a problem. In addition, listening tool between Social Networks and CRM systems is a mechanism to capture actual data from social media and propagates this information forward to the CRM. This tool should be able to filter noise (level of necessity for business process) from actual data that needs to be communicated to CRM.

Social CRM empowers patient/family to have the ability in controlling his own data. Once patient/family registers to have service from healthcare provider, it will enable them to have personalized e-health systems with Social CRM as the frontline of the system. The system will authorize for each patient then; the authorization and self-managed account/service are granted to access all applications and data offered by the systems. This authorization is expected to be in the long run since the information and contents continue to grow. Technical assistant is available through manual or health informatics officer (just like any other customer service in business/organization) who stand by online assisting patient/family in utilizing the system especially for the first timer. Furthermore, since all the information (medical records) can be accessed online anywhere and anytime, it will enable collaborative treatment from telemedicine.

Consider this scenario; while we go to physician for a diagnosis, sometimes there is tradeoff between time allocated to each patient and the comprehensiveness of diagnosing process. The long queue the patient waits for consultation enables the healthcare provider to allocate time wisely for each patient. Within the constraint of consultation time, the physician is able to conduct diagnosis efficiently and effectively. The system supports the customer service because it helps both the healthcare provider and patient in diagnosis activity. The physician will have complete information, knowledge, saving a lot of time to learn about patient history because patients participate in the detailing his medical records data through the system, and patient benefits from quality of diagnoses' time because his medical records are overviewed in full scene. In other words, it can provide better customer service to meet patient's expectation and improve the quality of consultation time. The physician is expected to have a comprehensive view of the patient's history before diagnosing or analyzing consulted symptoms. This can be achieved because physician will be able to observe the report of patient's medical history such as last medicine consumption, previous diagnoses, lab result, activities suggested by health educator etc. In addition, by empowering patients with medical data and personalized e-health, the healthcare needs to provide officer in duty (health educator/ health promoter) in order to interpret medical data or respond online query/consultation. The officer in duty is required to have an ability to interpret medical data and also familiar with the technical details of the systems.

Social CRM functionalities are composed from Marketing, Sales, and Customer Service which are operated to achieve business strategy of healthcare organization. For example, marketing strategy should accommodate social marketing to promote public health and commercial marketing to acquire more customers coming for services. Customer service will offer distinct value for each activity. The difference from the traditional CRM is that the state of empowering for self-managed data and authorization will encourage patients to willingly provide full data without hesitation. More data provided means more information available for the sake of analyzing for the interest of marketing, sales, and customer service.

The unique characteristics of value creation by adopting Social CRM is the ability to generate contents from both parties either from healthcare and also patients. Social CRM in healthcare provides value-added services to patients like openness of medical records, improving patient loyalty, creating better healthcare-patient communication, improving brand image and recognition and self-managed data which will improve health literacy, to reduce economic burden for society to the whole.

The framework above proposes and combines the concept of value chain and value shop. As discussed in previous section, the raw data arrives in one state, and leaves in another state. The patient enters ill and leaves well (hopefully). The activities of value chain are; arriving from registration, patient care, discharge, marketing, and service—producing data at respective state. However, the process of patient care is elaborated according to the value shop where value is created by mobilizing resources and activities to resolve a particular patient's problem. The five generic categories of primary value shop activities; Problem-finding and acquisition, choosing the overall approach to solving the problem execution, and control & evaluation

Furthermore, the framework accommodates Social/Commercial Marketing team. Social marketing is more prevalent in the government healthcare that operates as an agent of the public at large. Campaign of healthy life is example of Social Marketing for healthcare. It is not intended for commercial benefit for short term but it is beneficial for the community. On the other hand, commercial marketing is standard marketing strategy exist for any business entities. Both are acting in responds to the public demands like social networks, mailing list, blog, etc.

The other feature of the model is the robustness of systems because more applications/services will be added as characteristics of Web 2.0. Some of the features that available to the user are: updating personal data, Medical Records & History (medical treatment received, medicine consumption history, family illness history, genetic, medical imaging, x-ray, etc), Preference services, Transaction, Payment/Billing data, Activities, Personal Health Promotion and Education, Email, Appointment, Friend in networks, forums, chatting, etc.

## VI. FUTURE DIRECTIONS

The use of ICT in healthcare organizations has grown in the same pattern it is the growing within the larger industry landscape. The use of web technology, database management systems and network infrastructure are part of ICT initiative that will influence of healthcare practice and administration.

One such trend is the slowly adoption of e-health systems toward the use of EMR. The systems move patient information from paper to electronic file formats so they can be easily and effectively managed. However, an interesting fact to be noted is the tendency of people to know more and actively participate in the health promotion, prevention, and care together with the rights that will become a standard legislature guide the development of information systems that support these tendencies. Thus, the trend is towards more involvement of patient or citizen in receiving information, in decision making and in responsibility for own health. The prime feature of this trend to shift from healthcare-institution centred care to the citizen-centred care is the emphasis on continuity of care from prevention to rehabilitation. This vision can be achieved through shared care which builds on health telematics networks and services, linking hospitals, laboratories, pharmacies, primary care and social centres offering to individuals a 'virtual healthcare centre' with a single point of entry. Furthermore, this vision implies provision of health services to homes with innovative services such as personal health monitoring and support systems and user-friendly information systems for supporting health education and awareness [17].

There are various emerging tools and technologies in creating and managing HIS. Semantic Web is an extension of the World Wide Web which offers a united approach to knowledge management and information processing by using standards to represent machine-interpretable information. Thus semantic Web technology helps computers and people to work better together by giving the contents well-defined meanings. The semantic Web has also drawn attention in the medical research communities [7]. Semantic web services can support a service description language that can be used to enable an intelligent agent to behave more like a human user in locating suitable Web services. While Web services are software components or applications, which interact using open XML and Internet technologies. These technologies are used for expressing application logic and information, and for transporting information as messages [31]. They have significantly increased interest in Service oriented architectures (SOAs) [10]. The benefits of Web services include loose coupling, ease of integration and ease of accessibility.

Furthermore, Web 2.0 refers to Web-oriented applications and services that use the Internet as a platform, with its unique features, relying on its strengths rather than trying to make the Internet suit a particular application. With its promise of a more powerful, engaging and interactive user experience, Web 2.0 seems poised to revolutionize the way in which we interact with information resources. Web 2.0 is commonly associated with technologies such as weblogs (blogs), social bookmarking, wikis, podcasts, Really Simple Syndication (RSS) feeds (and other forms of many-to-many publishing), social software, and Web application programming interfaces (APIs) [20].

A HealthGrid [16] allows the gathering and sharing of many medical, health and clinical records/databanks maintained by disparate hospitals, health organizations, and drug companies. In other words, HealthGrid is an environment in which data of medical interest can be stored and made easily available to different actors in the healthcare system: physicians, allied professions, healthcare centres, administrators and, of course, patients and citizens in general. Also, the driving forces for the individual and commercial adoption of the VoIP are the significant cost savings, portability, and functionality that can be realized by switching some or all of their voice services to VoIP. Chen et al showed the integration of mobile health information system with VoIP technology in a wireless hospital [6].

Ubiquitous computing is a paradigm shift where technology becomes virtually invisible in our lives. The ubiquitous computing environment will make possible new forms of organizing, communicating, working and living. However, ubiquitous computing systems create new risks to security and privacy. To organize the u-healthcare infrastructure, it is necessary to establish a context-aware framework appropriate for the wearable computer or small-sized portable personal computer in ubiquitous environment [19]. The mobile health (m-health) can be defined as mobile communications network technologies for healthcare [18]. This concept represents the evolution of "traditional" e-health systems from desktop platforms and wired connections to the use of more compact devices and wireless connections in e-health systems.

The adoption of Social CRM to healthcare provider will improve customer service and prevent any dispute between healthcare organization and patient. We propose that Social CRM framework as alternative solution to the healthcare provider [1]. Our model takes into account the future directions in healthcare industry in related to the development of ICT. We are currently refining the model, developing system architecture and a prototype of CRM 2.0 for an e-health system.

## VI. CONCLUSIONS

Relationship management in healthcare industry is vitally important both for healthcare providers and patients (customers); therefore managing customer relationship is a key factor for healthcare providers to sustain their business in a competitive environment. With development of Web 2.0 that carries a network paradigm, we are developing a CRM 2.0 of Social CRM for healthcare industry. Our Social CRM model offers new outlook either from patient or healthcare. Some features offered are robustness of the systems, ingenuousness/openness of information sharing, and closeness of relationship between patient-healthcare and patient with others. A systems based on our model generates value in each activity to the customer to provide better service. It also empowers customers with the information accessibility. We believe with better service that generate values to customer will create customers' trust and loyalty that help customers and health providers sustaining relationship for mutual benefits.

# About the Authors

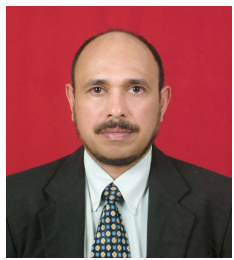

**Mohammad Nabil Almunawar** *Ir. IPB Indonesia, MSc UWO Canada, Ph.D. UNSW, Australia* is a senior lecturer at Faculty of Business, Economics and Policy Studies, Universiti of Brunei Darussalam (UBD), Brunei Darussalam. Dr Almunawar has published many papers in refereed journals as well as international conferences. He has many years teaching experiences and consultancies in the area computer and information systems. His overall research interest is application of IT in Management and e-commerce/e-business/e-government. He is also interested in object-oriented technology, multimedia information retrieval, health information systems and information security.

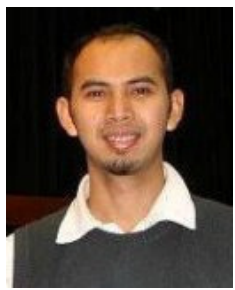

**Muhammad Anshari** is a Business Information Systems practitioner in Indonesia. He received his BMIS (Hons) from International Islamic University Malaysia, his Master of IT from James Cook University Australia, and currently, he is pursuing his PhD program at Universiti Brunei Darussalam. His professional experience started when he spent two years as an IT Business Analyst at Astra International Automotive Company, joined research collaboration on Halal Information Systems at Chulalongkorn University Thailand, and researcher at College of Computer and Information Science King Saud University, Saudi Arabia. His main research interest is on CRM, ERP, and SCM related to Health and Healthcare.